\documentclass[aps,prl,showpacs,twocolumn]{revtex4}
\usepackage{graphicx}

\def\ket{\rangle}
\def\<{\langle}
\def\>{\rangle}

\begin{document}
\title{quantum privacy amplification for quantum secure direct communication}
\author{}
\author{Fu-Guo Deng$^{1,3}$ and Gui Lu Long$^{1,2}$}
\address{$^{1}$Key Laboratory For Quantum Information and Measurements,
and Department of Physics, Tsinghua University, Beijing 100084,
P. R. China\\
 $^{2}$Center for Atomic and Molecular
NanoSciences, Tsinghua University, Beijing 100084, P. R. China\\
$3$ Key Laboratory of Beam Technology and Materials Modification
of MOE, and Institute of Low Energy Nuclear Physics, Beijing
Normal University, Beijing 100875, P. R. China}
\date{\today }

\begin{abstract}
Using quantum mechanics, secure direct communication between
distant parties can be performed. Over a noisy quantum channel,
quantum privacy amplification is a necessary step to ensure the
security of the message. In this paper, we present a quantum
privacy amplification scheme for quantum secure direct
communication using single photons.  The quantum privacy
amplification procedure contains  two control-not gates and a
Hadamard gate. After the unitary gate operations, a measurement is
performed and one photon is retained. The retained photon carries
the state information of the discarded photon, and hence reduces
the information leakage. The procedure can be performed
recursively so that the information leakage can be reduced to any
arbitrarily low level.
\end{abstract}

\pacs{03.67.Hk, 03.67.Dd, 03.67.-a}
\maketitle

Quantum secure direct communication (QSDC) allows two distant
parties to directly communicate securely.  It has attracted
attention recently
\cite{beige,bf,long1,long2,cai1,yan0,zhangzj,cai2}. In contrast,
quantum key distribution usually generates a string of random keys
\cite{bb84,ekert91}, and the secret message  is transmitted later
through a classical channel, usually using the Vernam one-time-pad
cipher-system\cite{vernam}. Recently, a QSDC protocol using batch
of single photons is proposed \cite{long2}. It has two distinct
features. First it uses single photons instead of entangled photon
pairs. Secondly, the transmission is operated in a batch by batch
manner. This feature is also reflected in the QSDC protocols in
Refs. \cite{long1,yan0,zhangzj}. The secret message is encoded and
released to the quantum channel in public only when the channel is
assured secure so that no secret message is leaked even though an
malicious eavesdropper may intercept the encoded qubits. Over a
noiseless quantum channel, the scheme is completely secure. In
practice, channel noise is inevitable, error correction and
privacy amplification must be used in order to reduce the
information leakage below the desired level.

The basic steps in the QSDC protocol in Ref.\cite{long2} contains
4 steps. First, Bob prepares a batch of $N$ single photons
randomly in one of four polarization states: $|0\ket=|+z\ket$,
$|1\ket=|-z\ket$, $|x\ket=(|0\ket+|1\ket)/\sqrt{2}$ and
$|-x\ket=(|0\ket-|1\ket)/\sqrt{2}$.  Bob sends this batch of
photons to Alice. Alice stores most of the single photons and
selects randomly a subset of single photons and performs
measurement using either the $\sigma_z$ or $\sigma_x$ basis. Alice
publishes the measuring-basis and the measured results of these
single photons. Upon these information, Bob determines the error
rate. If the error rate is higher than the threshold, the process
is aborted, and if the error rate is below the threshold, the
process continues and Alice encodes her message using two unitary
operations on the stored single photons and then sends them back
to Bob. Upon receiving these encoded photons, Bob makes
appropriate measurement and reads out the message.

Classical privacy amplification (CPA) \cite{bennettcpa} has been
used for QKD, for instance the BB84 protocol \cite{bb84}. Quantum
privacy amplification (QPA) \cite{bennettqpa,deutschqpa} has been
used for QKD using entangled quantum system, for instance for the
Ekert91 QKD scheme \cite{ekert91}. Without privacy amplification,
the error threshold in the proposed QSDC scheme in
ref.\cite{long2} will be very small. To allow the scheme to be
operable over a noisy channel, quantum privacy amplification has
to be used. In this paper, we present a quantum privacy
amplification scheme for QSDC (QSDC-QPA).  This QSDC-QPA can be
used not only for the QSDC protocol, but also for the BB84 QKD
protocol. It uses simple local gate operations and single particle
measurement, and these operations could be implemented by
technologies that is currently being developed.

The basic scientific problem for the proposed QSDC-QPA scheme can
be expressed as follows. Suppose Bob sends Alice a batch of single
photons, each photon is randomly prepared in one of the four
quantum states $|0\ket$, $|1\ket$, $(|0\ket+|1\ket)/2$ and
$(|0\ket-|1\ket)/2$, where $|0\ket$ and $|1\ket$ denote the
horizontal and vertical polarization state of a photon. Due to
channel noise and eavesdropping, an error bit rate $r$ is known
for the photon batch ($r$ is four times of the error bit rate
detected by Alice and Bob using random sampling, because
eavesdropper's interception causes only 25 percent of error). The
QSDC-QPA task is to condense a portion of photons from the batch
so that Eve's information about the condensed photons is reduced
to below a desired level.

\begin{figure}[!h]
\begin{center}
\includegraphics[width=8cm,angle=0]{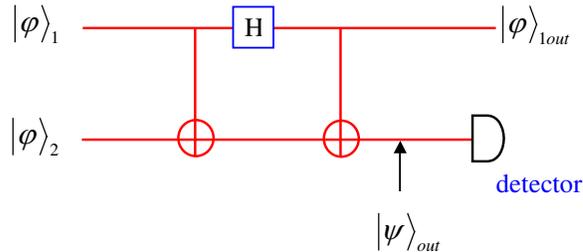} \label{f1}
\caption{ Quantum privacy amplification operation for two qubits.
}
\end{center}
\end{figure}

The basic operation of QSDC-QPA is shown in Fig.\ref{f1} for two
qubits. It consists of two control not (CNOT) gates and a Hadamard
(H) gate. Using the first qubit as control, two CNOT gates are
performed on the second qubit, or target qubit. Between these two
CNOT gates, a Hadamard gate is performed on the first qubit. Then
a measurement in the $\sigma_z$ basis is performed  on the second
qubit. Then the first qubit is the condensed qubit. We now explain
why the above procedure can reduce the information leakage.
Without loss of the generality, we assume the quantum states of
single photon 1 and 2 are written as
\begin{eqnarray}
\left\vert \varphi \right\rangle _{1}&=&a_{1}\left\vert
0\right\rangle +b_{1}\left\vert 1\right\rangle, \label{s1}\\
\left\vert \varphi \right\rangle _{2}&=&a_{2}\left\vert
0\right\rangle +b_{2}\left\vert 1\right\rangle, \label{s2}
\end{eqnarray}
where
\begin{equation}
\vert a_{1}\vert ^{2}+\vert b_{1}\vert ^{2}=\vert a_{2}\vert
^{2}+\vert b_{2}\vert ^{2}=1. \label{r1}
\end{equation}
 After the two CNOT gates and H-gate operations,
 the state of the joint system of single
photon 1 and 2 is changed to
\begin{eqnarray}
\left\vert \psi \right\rangle _{out} &=&\frac{1}{\sqrt{2}}%
\{(a_{1}a_{2}+b_{1}b_{2})\left\vert 0\right\rangle
_{1}+(a_{1}b_{2}-b_{1}a_{2})\left\vert 1\right\rangle
_{1}\}\left\vert
0\right\rangle _{2}  \nonumber \\
&+&\frac{1}{\sqrt{2}}\{(a_{1}a_{2}-b_{1}b_{2})\left\vert
1\right\rangle _{1}+(a_{1}b_{2}+b_{1}a_{2})\left\vert
0\right\rangle _{1}\}\left\vert 1\right\rangle _{2}.\nonumber\\
\label{s3}
\end{eqnarray}
After measuring the second qubit in the $\sigma_{z}$-basis, no
matter what the result, the state of the control qubit
$\vert\varphi\rangle_{1out}$ will contain the information of the
state of the original target qubit (qubit 2). Tables \ref{Table1}
and  \ref{Table2} give the output state of control qubit  after
the measurement on the target qubit with result $0$ and 1
respectively. It depends not only on the result of the measurement
on the target qubit, but also on the original states of the two
input single photons (photon 1 and 2).
\begin{table}
\begin{center}
\caption{The state of the output qubit when the result of the
second qubit measurement is $\vert 0\rangle$. $\varphi_1$ and
$\varphi_2$ are the original states of the  control   and target
qubit, respectively.}
\begin{tabular}{c|cccc}\hline
  & \multicolumn{4}{c}{$\varphi_{1}$}\\ \cline{2-5}
$\varphi_{2 }$& $\left\vert +z\right\rangle $ & $\left\vert -z\right\rangle $ & $%
\left\vert +x\right\rangle $ & $\left\vert -x\right\rangle$\\
\cline{2-5}
 $\left\vert +z\right\rangle $ & $\left\vert 0\right\rangle
$ & $\left\vert 1\right\rangle $ & $|-x\ket$  & $|+x\ket$ \\
$\left\vert -z\right\rangle $ & $\left\vert 1\right\rangle $ &
$\left\vert 0\right\rangle $ & $|+x\ket$ &
$|-x\ket$ \\
$\left\vert +x\right\rangle $ & $|+x\ket$ & $|-x\ket$ &
$\left\vert 0\right\rangle $ & $\left\vert 1\right\rangle $ \\
$\left\vert -x\right\rangle $ & $|-x\ket$ & $|+x\ket$ &
$\left\vert 1\right\rangle $ & $\left\vert 0\right\rangle $\\
\hline
\end{tabular}\label{Table1}
\end{center}
\end{table}
\begin{table}
\begin{center}
\caption{he state of the output qubit when the result of the
second qubit measurement is $\vert 1\rangle$. $\varphi_1$ and
$\varphi_2$ are the original states of the  control   and target
qubit, respectively.}
\begin{tabular}{c|cccc}\hline
 & \multicolumn{4}{c}{$\varphi_1$}\\ \cline{2-5}
$\varphi_2$ & $\left\vert +z\right\rangle $ & $\left\vert -z\right\rangle $ & $%
\left\vert +x\right\rangle $ & $\left\vert
-x\right\rangle$\\\hline $\left\vert +z\right\rangle $ &
$\left\vert 1\right\rangle $ & $\left\vert 0\right\rangle $ &
$|x\ket$ &
$|-x\ket$ \\
$\left\vert -z\right\rangle $ & $\left\vert 0\right\rangle $ &
$\left\vert 1\right\rangle $ & $|-x\ket$ &
$|+x\ket$ \\
$\left\vert +x\right\rangle $ & $|+x\ket$ & $|-x\ket$ &
$\left\vert 0\right\rangle $ & $\left\vert 1\right\rangle $ \\
$\left\vert -x\right\rangle $ & $|-x\ket$ & $|+x\ket$ &
$\left\vert 1\right\rangle $ & $\left\vert 0\right\rangle $\\
\hline
\end{tabular}\label{Table2}
\end{center}
\end{table}
Suppose that Eve knows the complete information of the  first
qubit. If the second photon is unknown to her, because the single
photons prepared by Bob is in one of the four states randomly with
$1/4$ probability each,  Eve's knowledge about the output state of
the control qubit after the quantum privacy amplification
operation becomes
\begin{eqnarray}
\rho &=&\frac{1}{4}\left(
\left\vert +z\right\rangle \left\langle +z\right\vert +%
\left\vert -z\right\rangle \left\langle -z\right\vert +%
\left\vert +x\right\rangle \left\langle +x\right\vert +\left\vert
-x\right\rangle \left\langle
-x\right\vert\right)\nonumber\\
 &=&\frac{1}{2}\left(
\begin{array}{cc}
1 & 0 \\
0 & 1%
\end{array}%
\right).
\end{eqnarray}
That is, Eve has no knowledge at all about the output state. But
for Bob who has prepared the original states of the two qubits, he
will know completely the output state when Alice tells him the
$\sigma_{2,z}$ measurement result.

If it happens that Eve has complete information about both qubits,
she will know the output state exactly just like Bob. However the
probability this can happen is only
\begin{eqnarray}
P_2 ={r^2}.
\end{eqnarray}
We can use the output qubit again as a control qubit and choose a
third qubit from the batch as the target qubit and perform QPA
operation on them. In this way, as more qubits are used in the QPA
process, Eve's information is reduced exponentially
\begin{eqnarray}
P_m={r^m},
\end{eqnarray}
where $m$ is the number of qubits that have been used in the GPA.
In this way, Alice can condense a portion of single photons from a
batch of $N$ photons with negligibly small information leakage.
Then Alice can perform unitary operations to encode her secret
message.

In summary, quantum privacy amplification on a batch of polarized
single photons can be done with quantum mechanics for secure
direct communication.

This work is supported the National Fundamental Research Program
Grant No. 001CB309308, China National Natural Science Foundation
Grant No. 60073009, 10325521, the Hang-Tian Science Fund, the
SRFDP program of Education Ministry of China.

\end{document}